\documentclass[10pt]{article} 

\usepackage{times}

\usepackage{geometry}              
\usepackage{microtype}
\usepackage{graphicx}
\usepackage{subfigure}
\usepackage{booktabs}
\usepackage{float}

\usepackage{hyperref}
\usepackage{xcolor} 

\definecolor{darkblue}{rgb}{0,0.08,0.50}
\hypersetup{
    colorlinks=true,        
    linkcolor=darkblue,         
    citecolor=darkblue,         
    filecolor=darkblue,      
    urlcolor=darkblue           
}

\usepackage[numbers]{natbib}  

\usepackage{fancyhdr} 

\newcommand{\documenttitle}{Adversarial Robustness in Distributed Quantum Machine Learning}

\fancypagestyle{plain}{%
  \fancyhf{} 
  \fancyfoot[C]{\thepage} 
}

\renewenvironment{abstract}{
	\centerline{\large \bf Abstract}
     \vspace{-0.1in}\begin{quote}
}
{\par\end{quote}\vskip 0.15in}

\usepackage{enumitem}
\usepackage{bm}
\usepackage{braket}
\usepackage{bbold}
\usepackage{qcircuit}

\usepackage{amsthm}
\usepackage{amsmath}
\usepackage{mathtools}
\usepackage{amssymb}

\theoremstyle{plain}
\newtheorem{theorem}{Theorem}[section]

\theoremstyle{definition}
\newtheorem{definition}[theorem]{Definition}

\theoremstyle{remark}

\usepackage{braket}

\begin{document}

\title{\documenttitle \thanks{
This is a preprint of the following planned book chapter: Pouya Kananian and Hans-Arno Jacobsen,  \textit{Adversarial Robustness in Distributed Quantum Machine Learning},  to be published in \textit{Quantum Robustness in Artificial Intelligence}, edited by Muhammad Usman,  to be published by Springer Nature as part of the \textit{Quantum Science and Technology} book series.  Reproduced with permission of Springer Nature. }}
\author{Pouya Kananian and Hans-Arno Jacobsen\\[0.5em]
Department of Electrical and Computer Engineering,  University of Toronto, Toronto, Canada\\[0.5em]
\texttt{pouya.kananian@mail.utoronto.ca, jacobsen@eecg.toronto.edu}}
\date{} 
\maketitle
\thispagestyle{plain}

\begin{abstract}
Studying adversarial robustness of quantum machine learning (QML) models is essential in order to understand their potential advantages over classical models and build trustworthy systems. Distributing QML models allows leveraging multiple quantum processors to overcome the limitations of individual devices and build scalable systems. However, this distribution can affect their adversarial robustness, potentially making them more vulnerable to new attacks. 
Key paradigms in distributed QML include federated learning, which, similar to classical models, involves training a shared model on local data and sending only the model updates, as well as circuit distribution methods inherent to quantum computing, such as circuit cutting and teleportation-based techniques. 
These quantum-specific methods enable the distributed execution of quantum circuits across multiple devices.
This work reviews the differences between these distribution methods, summarizes existing approaches on the adversarial robustness of QML models when distributed using each paradigm, and discusses open questions in this area.


\end{abstract}

\section{Introduction}
\label{sec:Intro}

Quantum machine learning (QML) is a rapidly developing area of research \cite{biamonte2017quantum,abbas2021power,liu2021rigorous,
cerezo2022challenges,larocca2024review}.
When evaluating quantum versus classical classifiers and studying the potential advantages of quantum models, adversarial robustness emerges as an important consideration \cite{west2023towards}.
As a result,  adversarial robustness in quantum machine learning has recently attracted significant attention \cite{lu2020quantum,liu2020vulnerability,weber2021optimal,du2021quantum,
liao2021robust,gong2022universal,
anil2024generating,dowling2024adversarial,kananian2025adversarial}.
By distributing QML models across multiple quantum processors,  we can overcome the limitations of individual devices and enable scalable quantum systems.  
However, this distribution introduces unique vulnerabilities,  which adversaries can exploit to launch more sophisticated and scalable attacks.  Grasping and addressing these challenges is critical to ensuring the security and reliability of distributed quantum machine learning systems, as well as to understanding their potential benefits compared to classical models.

Federated learning \cite{konevcny2016federated,konevcny2016federated,mcmahan2017communication} is one of the key paradigms in distributed machine learning.  
It allows multiple data owners to collaboratively train a shared model without sharing their local private data.
Quantum computing can be incorporated into federated learning via quantum data \cite{xia2021quantumfed,chehimi2022quantum}, quantum machine learning models \cite{huang2022quantum,kumar2023expressive}, or quantum communication \cite{li2021quantum,zhang2022federated,xu2023secure}.
A key benefit of adding quantum capabilities to federated learning is the ability to encode model parameters as quantum states, enabling secure and efficient communication through quantum channels \cite{li2021quantum,zhang2022federated,xu2022privacy,
kaewpuang2023adaptive,park2023hqk,ren2024variational,swaminathan2024distributed,tanbhir2025quantum}.
Furthermore, quantum models such as overparameterized variational classifiers may possess intrinsic robustness against adversarial attacks \cite{kumar2023expressive,heredge2024prospects,
gong2024enhancing,papadopoulos2025numerical}.
However, one of the reasons classical federated learning has attracted significant attention is the widespread availability of hardware resources—such as IoT devices, smartphones, and edge servers—that can be leveraged for training machine learning models in a decentralized manner. In contrast, quantum computing lacks such flexibility due to the scarcity and high cost of quantum hardware.
Therefore, one important paradigm in distributed quantum machine learning is how to distribute the execution of a single quantum model across multiple quantum processors, using either classical communication \cite{peng2020simulating,mitarai2021constructing} or quantum entanglement and teleportation protocols \cite{bennett1993teleporting,bouwmeester1997experimental,boschi1998experimental}.  This research direction is especially important given the current limitations of NISQ-era \cite{preskill2018quantum} quantum devices, such as limited qubit counts, short coherence times, and noisy operations, which constrain the scalability of quantum machine learning algorithms on individual devices.

While the robustness of quantum federated learning systems to adversarial and privacy-leaking attacks has been extensively studied
 \cite{li2021quantum,xia2021defending,zhang2022federated,
 kumar2023expressive,chu2023cryptoqfl,li2024privacy,
 chen2024robust,papadopoulos2025numerical,maouaki2025qfal}, adversarial robustness in quantum machine learning models when their circuits are distributed across multiple processors has only recently begun to receive attention \cite{kananian2025adversarial}.
This work provides an overview of quantum federated learning and circuit distribution methods in quantum computing,  outlining existing work on the adversarial robustness of quantum models distributed under each paradigm.
Although alternative distribution paradigms exist in quantum machine learning \cite{sheng2017distributed,yun2022quantum,park2023quantum,tuysuz2023classical,
cowlessur2024hybrid,kawase2024distributed,wu2024distributed,
hwang2024distributed,park2025addressing,chen2025distributed},  
we exclude them from the scope of this study.  These paradigms have received comparatively less attention in the context of adversarial robustness. 
Section  \ref{sec:back} provides background on classical and quantum federated learning, as well as circuit distribution methods in quantum computing using classical or quantum communication,  i.e., circuit cutting \cite{peng2020simulating,mitarai2021constructing,mitarai2021overhead} and teleportation-based methods.
In Section \ref{sec:adv-federated}, we review adversarial attacks in quantum  federated learning and potential defenses, while Section \ref{sec:adv-cut} focuses on the adversarial robustness of quantum models with partitioned circuits.

\section{Background}
\label{sec:back}

\subsection{Federated Learning}
\label{subsec:federated}

Federated learning (FL) \cite{konevcny2016federated,
konevcny2016federated,mcmahan2017communication} 
is a machine learning approach that enables multiple data owners to collaboratively train a shared model while keeping their local data private.  Moreover,  the shared model should achieve accuracy close to what would have been obtained if all the data had been centrally aggregated and used to train a single model \cite{yang2019federated}.  Instead of sending data to a central server, 
clients can train models locally and only communicate model parameters.
Federated Learning systems often employ diverse hardware and use datasets that are typically non-IID and imbalanced in terms of size, diversity, and quality
\cite{li2020multi}.  

Federated learning is typically classified into three primary types: vertical,  horizontal,  and a variant known as federated transfer learning \cite{wen2023survey,saha2024multifaceted,zhang2024survey}.  
In horizontal (or homogeneous) federated learning, clients share the same types of features in their data, though the actual data samples differ among them \cite{haddadpour2019convergence}.
Clients keep their private data local, exchanging only global and local model parameters with the server during communication.  
The server trains the global model by collecting and aggregating model parameters or gradients from the clients.
In contrast,  vertical (or heterogeneous) federated learning refers to situations where clients share the same data samples but have different sets of features \cite{liu2020secure,yu2020heterogeneous}.
Typically, one client is assumed to possess all the labeled data and is termed the guest or active client, whereas the remaining clients, which do not have labels, are called host or passive clients \cite{aledhari2020federated, cheng2021secureboost}.
 In contrast to horizontal federated learning, which produces a unified global model, the vertical setup results in distinct local models for each party, requiring collaboration among clients to perform inference \cite{liu2024vertical}.
 In federated transfer learning, both the features and samples vary between datasets, albeit with some overlap \cite{liu2024vertical}.

To build machine learning systems that are both scalable and capable of leveraging rich, diverse data sources, it is essential to move beyond centralized training and embrace distributed approaches like federated learning. In many real-world scenarios, data is naturally distributed across a wide range of devices or organizations, and collecting it centrally is impractical due to privacy concerns, bandwidth limitations, or regulatory constraints. 
However, like other distributed systems, federated learning introduces additional attack surfaces
and potential vulnerabilities related to security, privacy, and fairness \cite{zhang2024survey,wei2025trustworthy}.
Although private data is not shared in federated learning, the exchanged models and gradients can still expose sensitive information, and FL systems remain vulnerable to inference and poisoning attacks \cite{kairouz2021advances,qammar2022federated,
zhang2022survey,almutairi2023federated,
xie2024survey,zhang2024survey}.  
As a result, there is a growing need to develop trustworthy federated learning systems that ensure robust protections while maintaining the benefits of decentralized learning \cite{zhang2024survey}.
Trustworthy federated learning systems should ensure privacy by safeguarding sensitive data from exposure, maintain robustness even under adversarial conditions, uphold fairness, and support explainability—both through transparent, interpretable system design and through external mechanisms that help elucidate the model’s decisions \cite{zhang2024survey}.
 Therefore, to build such systems, it is imperative to study adversarial robustness in federated learning, along with the closely related challenge of privacy preservation.

\subsection{Quantum Federated Learning}
\label{subsec:quantum-federated}

Quantum computing can be integrated into federated learning through the use of quantum data \cite{xia2021quantumfed,chehimi2022quantum},  quantum machine learning models \cite{xia2021quantumfed,huang2022quantum,chehimi2022quantum,kumar2023expressive},
 or quantum communication \cite{li2021quantum,zhang2022federated,xu2023secure,narottama2023federated,kannan2024quantum}. 
Numerous frameworks for quantum federated learning have been proposed \cite{chen2021federated,xia2021quantumfed,
huang2022quantum,chehimi2022quantum,yun2022slimmable,
qi2023optimizing,gurung2023decentralized,qu2023dtqfl,
lusnig2024hybrid,innan2024fedqnn,qu2024quantum,swaminathan2024distributed,gurung2025quantum},
  as well as quantum-inspired approaches \cite{yamany2021oqfl,subramanian2024hybrid}.
For instance,  Chen et al. \cite{chen2021federated} pioneered a quantum federated learning framework integrating hybrid quantum-classical networks, where classical neural networks extract features that are subsequently processed by quantum circuits.
Xia et al. \cite{xia2021quantumfed} propose a framework in which multiple quantum nodes train a quantum neural network \cite{kak1995quantum,cong2019quantum,pesah2021absence,du2021quantum} using their local quantum data.  
Additionally, there exist frameworks that allow quantum model training across classical clients \cite{song2024quantum}.  
In Song et al.'s work \cite{song2024quantum},  
shadow tomography \cite{huang2020predicting,huang2021efficient} 
is used by the server to generate a classical approximation of the quantum model, allowing clients to calculate local gradients on their own data.
To learn more about developments in this area, readers can refer to several available surveys \cite{ren2023towards,gurung2023quantum,chehimi2023foundations,
saha2024multifaceted,
qiao2024transitioning,mathur2025federated}.

One advantage of incorporating quantum capabilities into federated learning is the ability to encode classical data using a logarithmic number of qubits. This allows for inference and training of certain machine learning models via gradient descent with exponentially lower communication costs compared to frameworks relying on classical communication \cite{gilboa2024exponential}.
Moreover, encoding model parameters into quantum states facilitates secure and efficient communication over quantum channels \cite{li2021quantum,zhang2022federated,xu2022privacy,
kaewpuang2023adaptive,park2023hqk,ren2024variational,swaminathan2024distributed,tanbhir2025quantum}, employing methods such as quantum key distribution \cite{bennett1984quantum,shor2000simple,bennett2014quantum,mehic2020quantum}, quantum secret sharing \cite{hillery1999quantum}, and blind quantum computing \cite{broadbent2009universal,polacchi2023multi}.
Leveraging quantum mechanical principles like the no-cloning theorem \cite{wootters1982single,dieks1982communication,milonni1982photons,barnum1996noncommuting}, these approaches offer a secure alternative that lessens reliance on computationally intensive encryption \cite{sheng2017distributed,  zhang2022federated}. 
Furthermore, certain quantum machine learning models, such as overparameterized variational classifiers, may offer inherent resilience to adversarial attacks \cite{kumar2023expressive,heredge2024prospects,
gong2024enhancing,papadopoulos2025numerical}.
The potential advantages of quantum federated learning frameworks in terms of privacy preservation and resilience to adversarial attacks are explored further in Section \ref{subsec:adv-federated-priv-leak}.

\subsection{Circuit Cutting}
\label{subsec:circ-cut}

Noisy Intermediate-Scale Quantum (NISQ) \cite{preskill2018quantum} devices represent the current generation of quantum computers, characterized by having tens to a few hundred qubits without full error correction capabilities. These devices mark a significant step toward practical quantum computing but are still limited by short coherence times, gate errors, and noise that degrade computational accuracy. 
Due to the limited qubit capacity of NISQ-era devices, a major challenge is that some quantum circuits exceed the size that current quantum processors can handle.  
To address this limitation, various methods  \cite{bravyi2016trading, peng2020simulating,yuan2021quantum,mitarai2021constructing,mitarai2021overhead,eddins2022doubling,fujii2022deep, wiersema2022circuit} have been proposed that leverage classical processing to enable execution on these constrained devices.
A significant category of these approaches is circuit cutting \cite{lowe2023fast}—also known as circuit knitting \cite{piveteau2023circuit}, circuit decomposition, or circuit fragmentation \cite{ufrecht2023cutting}— 
 with most methods in this category relying on
quasiprobability simulation,  a core method also widely applied in  quantum errors mitigation 
 \cite{temme2017error,endo2018practical,piveteau2022quasiprobability} and classical simulation of quantum systems \cite{pashayan2015estimating, howard2017application,seddon2019quantifying,seddon2021quantifying}.
Circuit cutting entails dividing quantum circuits into smaller subcircuits.  After executing these subcircuits,  their outcomes could be combined using classical post-processing to simulate quantum circuits that need more qubits than are available on a specific quantum processor.
Circuit cutting techniques generally fall into two categories.
In wire cutting \cite{peng2020simulating,uchehara2022rotation,lowe2023fast,pednault2023alternative,brenner2023optimal,harada2023doubly,harrow2024optimal,li2024efficient},
the quantum identity channel is decomposed into a linear combination of measure-and-prepare channels, 
while gate cutting \cite{mitarai2021constructing,mitarai2021overhead,
piveteau2023circuit, schmitt2023cutting, ufrecht2023cutting, ufrecht2024optimal,harrow2024optimal} involves breaking down a non-local channel using a sum of tensor products of local
channels. 

\subsubsection{Quasiprobability Decomposition}
\label{subsec:quasi}

Quasiprobability simulation \cite{pashayan2015estimating,temme2017error,endo2018practical,
mitarai2021constructing,piveteau2023circuit} 
 and circuit cutting \cite{peng2020simulating,mitarai2021constructing,mitarai2021overhead} have mostly been explored in the context of circuits where the output is the expectation value of an observable $O$.  The goal of these circuits is to estimate the expected value $\langle O \rangle = \text{Tr}(O \mathcal{E}(\sigma))$,  where $\mathcal{E}$ represents the quantum channel realized by the circuit, and $\sigma$ is the input quantum state.
Quasiprobability simulation involves replacing a quantum channel $\mathcal{V}$ with a linear combination of implementable channels $\{\mathcal{E}_i\}$,  according to the following decomposition.
\begin{align}
\mathcal{V}(\rho) = \sum_i c_i \mathcal{E}_i(\rho),
\label{equ:quasiprob}
\end{align}
where $c_i \in \mathbb{R}$ and $\rho$ is a quantum state.  The term \textit{quasiprobability} comes from the fact that the coefficients can be negative; therefore,  they are not true probabilities.  Rewriting the decomposition (\ref{equ:quasiprob}) as follows allows us to obtain the expectation value of the circuit using Monte Carlo sampling.
\begin{align*}
\mathcal{V}(\rho) = \sum_i p_i \mathcal{E}_i(\rho) .  \text{sign}(c_i)\left(
\sum_i \vert c_i \vert\right),
\end{align*}
where $p_i := \vert c_i \vert / (\sum_i \vert c_i \vert)$.
Using Monte Carlo sampling,  each shot of the circuit samples an index $i$ randomly according to the probability distribution $\{p_i\}$,  and replaces the channel $\mathcal{V}$ with the corresponding channel $\mathcal{E}_i$.  The measurement outcome from this shot is then multiplied by the weight $\text{sign}(c_i)(\sum_i \vert c_i \vert)$,  where $c_i$ is the quasiprobability coefficient associated with $\mathcal{E}_i$.  This process is repeated over many shots, and the final estimate of the observable's expectation value is obtained by averaging the weighted measurement outcomes.
In this approach,   measurement outcomes are reweighted by a factor proportional to $\kappa = \sum_i \vert c_i \vert$,  which introduces a sampling overhead.  Specifically, using Hoeffding’s inequality,  to estimate the expectation value of an observable to within an additive error $\epsilon$ with probability at least $1 - \delta$,   the number of required circuit executions (shots) scales as 
\begin{align*}
N = a.\left(\sum_i \vert c_i \vert\right)^2 \epsilon^{-2} \ln{\frac{1}{2\delta}} = \mathcal{O}(\kappa^2),
\end{align*}
for some constant $a$.

\subsubsection{Gate Cutting}
\label{subsec:gate-cut}

\begin{figure}[t]
\includegraphics[width=\textwidth]{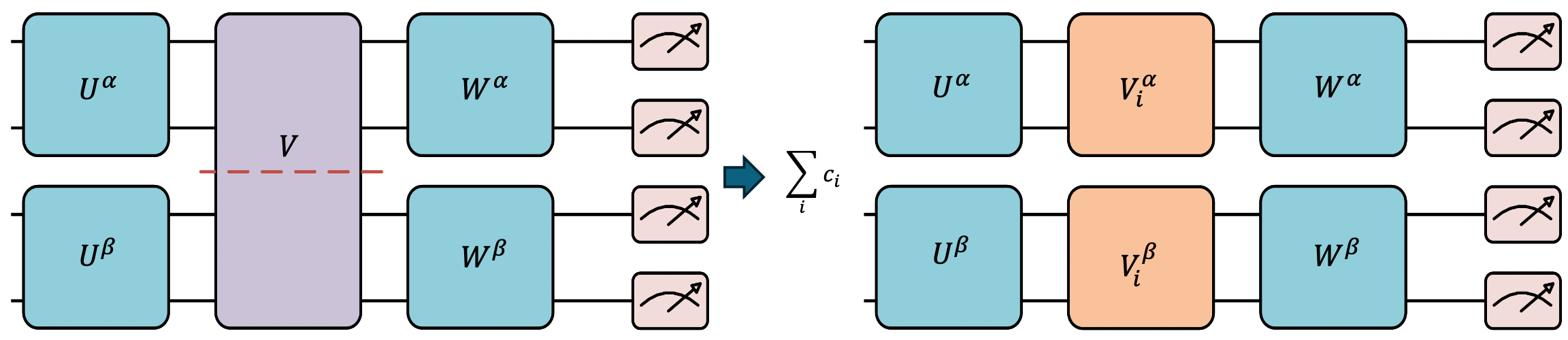}
\caption{By applying gate cutting to the quantum circuit on the left, we can quasiprobabilistically decompose gate $V$
using Equation (\ref{equ:gate-cut}), replacing it with a linear combination of local gates.  
To simulate the original circuit on the left, one can execute the corresponding subcircuits on the right and reconstruct the result using classical post-processing.}
\label{fig:gate-cut}       
\end{figure}

Similar to Section \ref{subsec:quasi},  for gate cutting and wire cutting, consider a circuit whose goal is to estimate the expectation value of an observable.  Suppose the qubits are divided into two partitions, $\alpha$ and $\beta$.  
Consider a multi-qubit $V$ that acts across the two partitions, and let 
$\mathcal{V}(.) = V(.)V^\dagger$ denote its corresponding unitary channel.
The objective of gate cutting \cite{mitarai2021constructing,mitarai2021overhead} is to express this unitary channel as a decomposition,
\begin{align}
\mathcal{V}(\rho) = \sum_i c_i \left( \mathcal{V}_i^\alpha (\rho^\alpha) \otimes \mathcal{V}_i^\beta (\rho^\beta)  \right),
\label{equ:gate-cut}
\end{align}
where $\mathcal{V}_i^\alpha$ and $\mathcal{V}_i^\beta$ are local channels acting on partitions $\alpha$ and $\beta$,  respectively,  and 
$\rho = \rho^\alpha \otimes \rho^\beta$,  with $\rho^\alpha$ and $\rho^\beta$ denoting the marginal states of $\rho$ corresponding to these two partitions. 
Figure \ref{fig:gate-cut} illustrates the application of gate cutting to a simple quantum circuit.
Let $\mathcal{U}$ and $\mathcal{W}$ denote the unitary channels corresponding to $U^\alpha \otimes U^\beta$ and $W^\alpha \otimes W^\beta$, respectively.  The expectation value of the observable $O$ is given by:
\begin{align*}
\langle O \rangle = \text{Tr}(O \mathcal{W} \circ \mathcal{V} \circ \mathcal {U}( \sigma)),
\end{align*}
where $\mathcal{U}(.) = (U^\alpha \otimes U^\beta)(.)(U^\alpha \otimes U^\beta)^\dagger$, $\mathcal{W}(.) = (W^\alpha \otimes W^\beta)(.)(W^\alpha \otimes W^\beta)^\dagger$,  and $\sigma$ denotes the input quantum state.
After applying gate cutting, the expectation value can be expressed as:
\begin{align*}
\langle O \rangle &= \sum_i c_i \text{Tr}((O^\alpha \otimes O^\beta)  \left(
(\mathcal{W}^\alpha \circ \mathcal{V}_i^\alpha \circ \mathcal{U}^\alpha)(\sigma^\alpha)
\otimes (\mathcal{W}^\beta \circ \mathcal{V}_i^\beta \circ \mathcal{U}^\beta)(\sigma^\beta)
\right) \nonumber \\
&= \sum_i c_i \text{Tr}\left(O^\alpha (\mathcal{W}^\alpha \circ \mathcal{V}_i^\alpha \circ \mathcal{U}^\alpha)(\sigma^\alpha)\right)
\text{Tr}\left(O^\beta (\mathcal{W}^\beta \circ \mathcal{V}_i^\beta \circ \mathcal{U}^\beta)(\sigma^\beta)\right)  \nonumber \\
&= \sum_i c_i \langle O^\alpha \rangle_i \langle O^\beta \rangle_i,
\end{align*}
where $O = O^\alpha \otimes O^\beta$,  with $O^\alpha$ and $O^\beta$ acting on partitions $\alpha$ and $\beta$,  respectively.  Here, 
$\sigma = \sigma^\alpha \otimes \sigma^\beta$,
$\mathcal{U} = \mathcal{U}^\alpha \otimes \mathcal{U}^\beta$,
 and  $\mathcal{W} = \mathcal{W}^\alpha \otimes \mathcal{W}^\beta$,  where 
$\mathcal{U}^\alpha(.) = U^\alpha(.)U^{\alpha\dagger}(.)$,  $\mathcal{U}^\beta(.) = U^\beta(.)U^{\beta\dagger}(.)$,  
$\mathcal{W}^\alpha(.) = W^\alpha(.)W^{\alpha\dagger}(.)$,  and $\mathcal{W}^\beta(.) = W^\beta(.)W^{\beta\dagger}(.)$.  We employ $\langle O^\alpha \rangle_i$ and $\langle O^\beta \rangle_i$ to denote the expectation values of $O^\alpha$ and $O^\beta$,  respectively,  when $\mathcal{V}$ is replaced by $\mathcal{V}_i^\alpha$ and $\mathcal{V}_i^\beta$.

As discussed in Section \ref{subsec:quasi},  Monte Carlo sampling can be used to estimate the expectation value.
The sampling overhead associated with gate cutting scales as $\mathcal{O}(\kappa^2)$,  where $\kappa = \sum_i \vert c_i \vert$ \cite{pashayan2015estimating,piveteau2022quasiprobability,ufrecht2023cutting}.   If $m$ gates are cut in the circuit,  the sampling overhead scales exponentially with $m$.  When these gates are cut separately by applying decomposition (\ref{equ:gate-cut}) to each one, the sampling overhead becomes $\mathcal{O}(\kappa^{2m})$.  However, more efficient methods for jointly cutting multiple gates have been proposed \cite{piveteau2023circuit, schmitt2023cutting, ufrecht2023cutting, ufrecht2024optimal,harrow2024optimal},  which reduce the sampling overhead—though it still scales exponentially with $m$.

\subsubsection{Wire Cutting}
\label{subsec:wire-cut}

\begin{figure}[t]
\includegraphics[width=\textwidth]{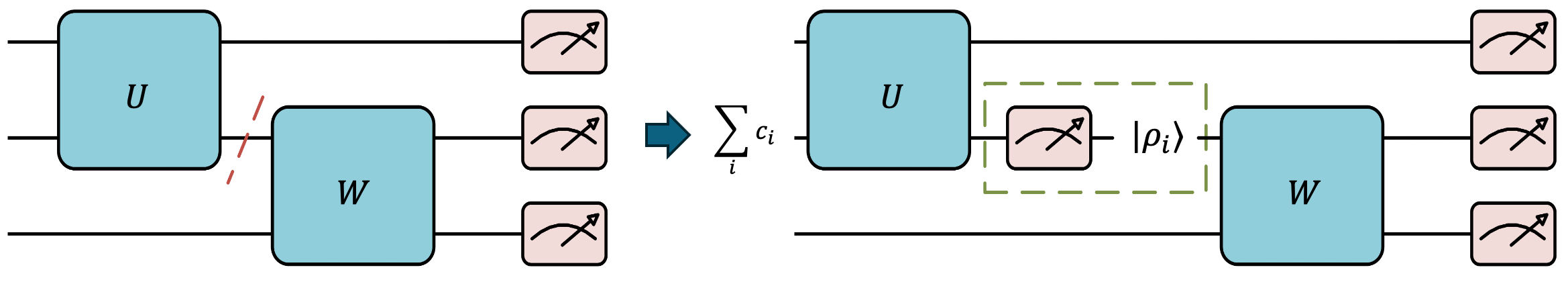}
\caption{
This figure depicts a simple quantum circuit that has been partitioned using wire cutting by applying Equation (\ref{equ:wire-cut})  to the wire connecting gates $U$  and $W$.   As with gate cutting, the original circuit can be simulated by executing the subcircuits on the right and combining their outcomes via classical post-processing.}
\label{fig:wire-cut}      
\end{figure}

Wire cutting was originally introduced by Peng et al.  \cite{peng2020simulating}.
The objective of wire cutting is to replace a wire—i.e., an identity channel—with a linear combination of measurement and state preparation operations.   The single-qubit identity channel can be decomposed as follows.
\begin{align}
\mathcal{I}(\rho) = \sum_i c_i \rho_i \text{Tr}(O_i \rho),
\label{equ:wire-cut}
\end{align} 
where $c_i \in \mathbb{R}$,  $\rho$ and $\rho_i$ are density matrices,  and $O_i$ denotes an observable corresponding to a measurement.
After applying wire cutting to the simple circuit shown in Figure \ref{fig:wire-cut}, the expectation value $\langle O \rangle$ can be expressed as follows. 
\begin{align*}
\langle O \rangle 
&= 
\sum_i c_i  \text{Tr}\left( (O_i \otimes O) ( \mathcal{U} (\sigma^\alpha) \otimes \mathcal{W}(\rho_i \otimes \sigma^\beta)) \right) \nonumber \\
&= 
\sum_i c_i  \text{Tr}\left( (O_i \otimes (O^\alpha \otimes O^\beta)) ( \mathcal{U} (\sigma^\alpha) \otimes \mathcal{W}(\rho_i \otimes \sigma^\beta)) \right) \nonumber \\
&= 
\sum_i c_i \text{Tr} \Big( (O_i \otimes O^\alpha) \mathcal{U} (\sigma^\alpha) \Big)
\text{Tr}\left( O^\beta \mathcal{W}(\rho_i \otimes \sigma^\beta))  \right) \nonumber \\
&= 
\sum_i c_i \langle O_i \otimes  O^\alpha \rangle \langle O^\beta \rangle_i,
\end{align*}
where $\mathcal{U}(.) = U(.)U^\dagger$ and $\mathcal{W}(.) = W(.)W^\dagger$ represent unitary channels,  and $\sigma = \sigma^\alpha \otimes \sigma^\beta$,  with $\sigma^\alpha$ and $\sigma^\beta$ denoting the 
marginal states of $\sigma$ associated with the top and bottom subcircuits,  respectively.
Similarly,  the observable is factorized as $O = O^\alpha \otimes O^\beta$,  where $O^\alpha$ and $O^\beta$ correspond to these two subcircuits.
Here,  $\langle O_i \otimes O^\alpha \rangle$ and $\langle O^\beta \rangle_i$ represent the expectation values of $O_i \otimes O^\alpha $ and $O^\beta$,  when 
the wire 
is replaced by the measure-and-prepare channel corresponding to $O_i$ and $\rho_i$.

The following is an example of a set of observables,  quantum states,  and real coefficients $\{O_i, \rho_i,  c_i \}_{i=0}^8$ that satisfies Equation (\ref{equ:wire-cut}) \cite{peng2020simulating}.
\begin{align}
\begin{aligned}
\{O_1 &= I, & \rho_1 &= |0\rangle\langle 0|, & c_1 &= +\frac{1}{2}, \\
O_2 &= I, & \rho_2 &= |1\rangle\langle 1|, & c_2 &= +\frac{1}{2}, \\
O_3 &= X, & \rho_3 &= |+\rangle\langle +|, & c_3 &= +\frac{1}{2}, \\
O_4 &= X, & \rho_4 &= |-\rangle\langle -|, & c_4 &= -\frac{1}{2}, \\
O_5 &= Y, & \rho_5 &= |+i\rangle\langle +i|, & c_5 &= +\frac{1}{2}, \\
O_6 &= Y, & \rho_6 &= |-i\rangle\langle -i|, & c_6 &= -\frac{1}{2}, \\
O_7 &= Z, & \rho_7 &= |0\rangle\langle 0|, & c_7 &= +\frac{1}{2}, \\
O_8 &= Z, & \rho_8 &= |1\rangle\langle 1|, & c_8 &= -\frac{1}{2}\},
\end{aligned}
\label{equ-set-satisfying-wire-cut}
\end{align}
where  $I, X,Y$ and $Z$ denote single-qubit Pauli matrices,  with
$|\pm\rangle$ and $|\pm i\rangle$ representing  $(|0\rangle \pm |1\rangle)/\sqrt{2}$ and $(|0\rangle \pm i|1\rangle)/\sqrt{2}$,  respectively.
This set can be obtained because any $2 \times 2$ density operator $\rho$ can be expanded using the normalized Pauli matrices, which form an orthonormal basis:
\begin{align}
\rho = 
\sum_{\tilde{B} \in \{ I,  X,  Y,  Z \} / \sqrt{2}} \text{Tr} (\tilde{B} \rho) \tilde{B}
= \frac{1}{2} \sum_{B \in \{ I,  X,  Y,  Z \}} \text{Tr} (B \rho) B.
\label{equ-expand-norm-pauli}
\end{align}
Expanding each Pauli matrix in its eigenbasis in Equation (\ref{equ-expand-norm-pauli}) reveals how the set (\ref{equ-set-satisfying-wire-cut}) satisfies Equation (\ref{equ:wire-cut}) \cite{peng2020simulating}.  
Decomposition (\ref{equ:wire-cut}) can be generalized to account for cutting $m$ parallel wires \cite{harada2023doubly}:
\begin{align}
\mathcal{I}^{\otimes m}(\rho) = \frac{1}{2^m} \sum_{P \in \{ I,X,Y,Z \}^{\otimes m}} \text{Tr} (P \rho) P,
\label{equ:parallel-cut}
\end{align}
where $P$ represents an $m-$qubit Pauli string.  
Since each Pauli matrix in the Pauli strings can be expanded in its eigenbasis,  
each term $\text{Tr} (P (.)) P$ reflects a measurement-preparation process 
in which the expectation value of $P$ is measured,  followed by the preparation of its eigenstates 
as input to the following subcircuit. 

Similar to gate cutting,  the sampling overhead associated with wire cutting  scales as $\mathcal{O}(\kappa^{2m})$ when $m$ wires are independently cut using decomposition (\ref{equ:wire-cut}),  where $\kappa = \sum_i \vert c_i \vert$.  When the values from set (\ref{equ-set-satisfying-wire-cut}) are substituted into decomposition (\ref{equ:wire-cut}),  we obtain $\kappa = \sum_{i=1}^{8} 1/2 = 4$,  resulting in the sampling overhead of $\mathcal{O}(\kappa^{2m}) = \mathcal{O}(16^{m})$.  Using decomposition (\ref{equ:parallel-cut}) to cut $m$ parallel wires results in a similar sampling overhead as decomposition (\ref{equ:wire-cut}).  However,  like gate cutting,  more efficient methods have been proposed for the joint cutting of $m$ wires \cite{brenner2023optimal,lowe2023fast,pednault2023alternative,harada2023doubly,harrow2024optimal}.
For example,  
Harada et al. \cite{harada2023doubly} propose a decomposition that not only achieves optimal sampling overhead for cutting $m$ parallel wires but also minimizes the number of quantum channels required for this task.  More efficient approaches exist for cutting non-parallel wires; however, they require assistance from ancilla bits \cite{brenner2023optimal}.  The decomposition proposed by Harada et al.  \cite{harada2023doubly} achieves a sampling overhead of $\mathcal{O}((2^{m+1} - 1)^2)$ for cutting $m$ parallel wires and $\mathcal{O}(9^m)$  for arbitrarily located wires.
In contrast, the decomposition introduced by Brenner et al.  \cite{brenner2023optimal} achieves a sampling overhead of $\mathcal{O}((2^{m+1} - 1)^2)$  for cutting non-parallel wires, which has been shown to be optimal 
when wire cutting can utilize arbitrary local operations and classical communication (LOCC) \cite{brenner2023optimal}.

\subsubsection{Circuit Cutting and Quantum Machine Learning}
\label{subsec:cut-qml}

Circuit cutting has attracted growing interest in recent years \cite{perlin2021quantum,ayanzadeh2023frozenqubits,
nagai2023quantum,tomesh2023divide,perez2023shallow,
gentinetta2024overhead,barral2024review},  with research focusing on a wide range of areas.  Notable examples include approximate circuit reconstruction \cite{chen2022approximate,lian2023fast,chen2023efficient,chen2023online}, 
intelligent qubit assignment across processors 
\cite{brandhofer2023optimal},  
finding optimal cut locations \cite{tang2021cutqc,tang2022scaleqc,smith2023clifford},  
distributed scheduling of circuit partitions \cite{bhoumik2023distributed,seitz2024multithreaded}, 
and the intersection of circuit cutting and quantum error mitigation
 \cite{majumdar2022error,liu2022classical, li2023enhancing,karim2024low}.
Circuit cutting can be used to distribute the execution of a quantum machine learning circuit across multiple quantum processors \cite{pira2023invitation,seitz2024multithreaded}.  
Moreover,  it can be integrated with federated learning systems to distribute the training of local models across multiple participants, enhancing the suitability of such systems for noisy, resource-constrained quantum processors \cite{sahu2024nac}. 
However, implementing circuit cutting in quantum machine learning presents notable challenges \cite{marshall2023high,guala2023practical,kananian2025adversarial}.

\begin{figure}[tb]
    \centering
    \includegraphics[width=0.8\textwidth]{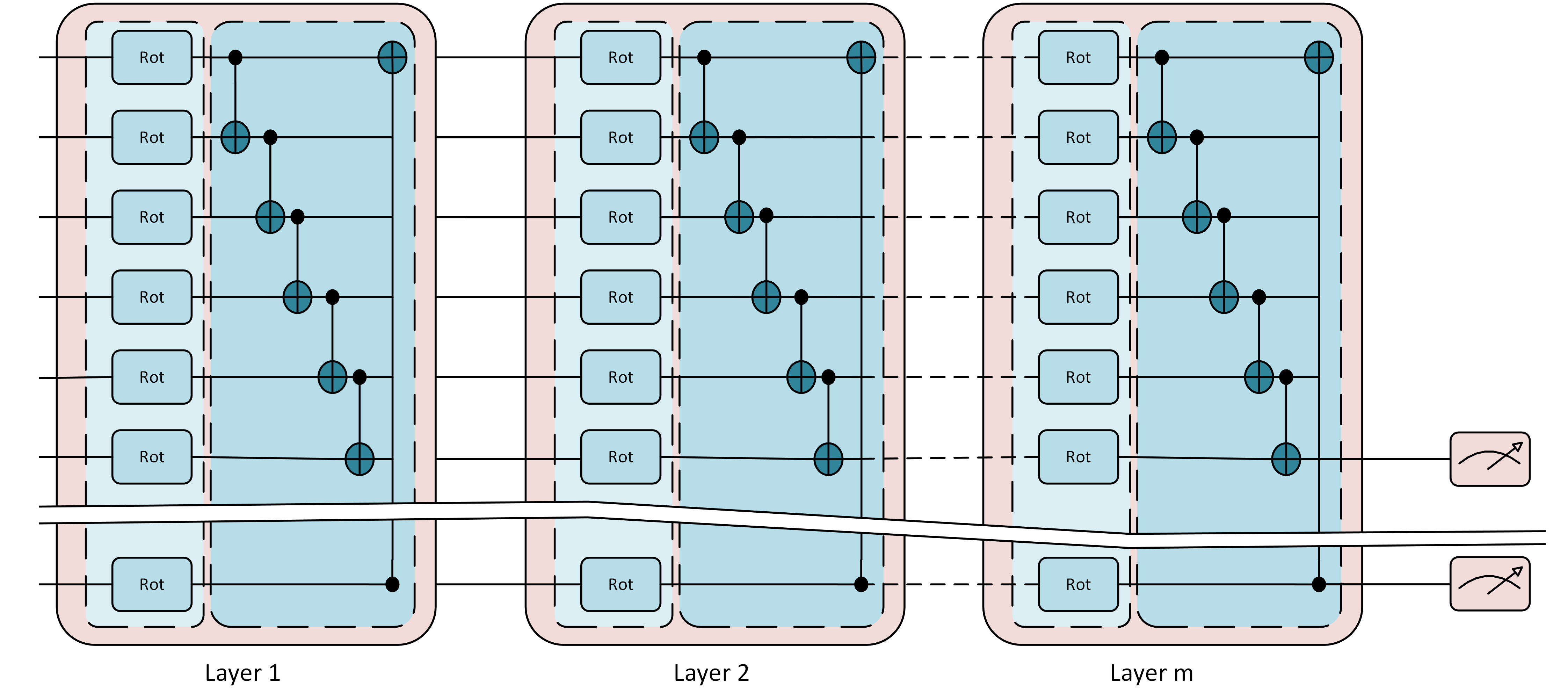}
    \caption{A strongly entangled 
    quantum circuit with $m$ layers.  In this circuit, each layer comprises a series of rotation gates followed by an entangling layer that fully entangles all the qubits.}
    \label{fig:strong-ent}
\end{figure}

The exponential sampling overhead associated with circuit cutting becomes especially problematic when applied to 
strongly entangled 
ansätze (see Fig. ~\ref{fig:strong-ent}),  which are commonly used in quantum machine learning \cite{kananian2025adversarial}.   All qubits are interconnected in this ansätze; therefore,  for an $n$-qubit circuit, for instance, in wire cutting, at least $n$ wires need to be cut to obtain two separate subcircuits.  For these ansätze, we can turn to approximation techniques to reduce the cost of circuit reconstruction at the expense of some accuracy \cite{marshall2023high}.

\begin{figure}[tb]
    \centering
    \includegraphics[width=0.5\textwidth]{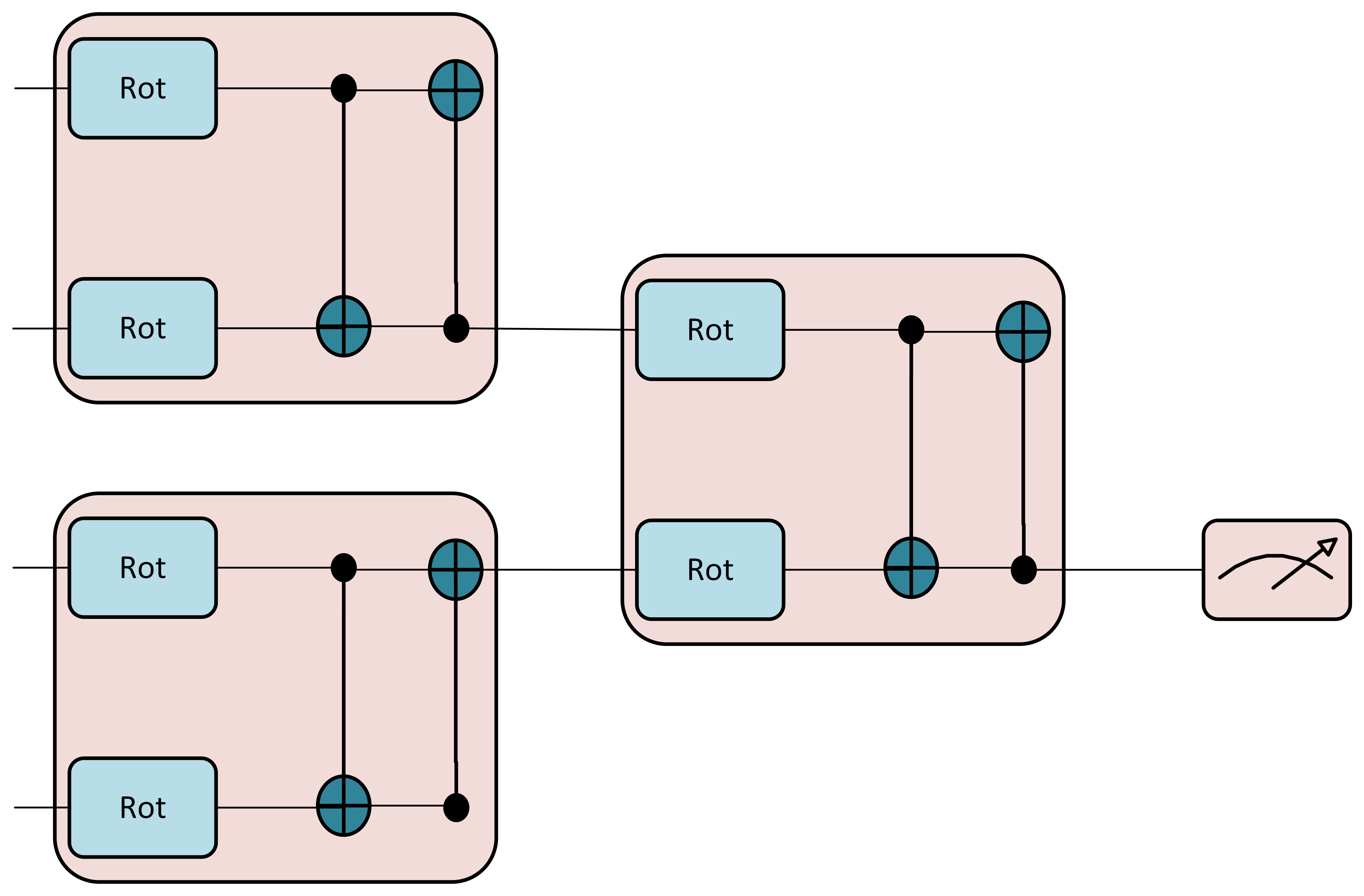}
    \caption{A simple tree tensor quantum circuit.}
    \label{fig:tree-tensor}
\end{figure}

Ansätze based on tree tensor networks \cite{shi2006classical,tagliacozzo2009simulation} (see Fig.~\ref{fig:tree-tensor}) are more compatible with integration with circuit cutting \cite{guala2023practical}.  When cutting them,  each tensor block could correspond to a subcircuit.  
This allows them to be executed on a processor with fewer qubits, while the number of circuit evaluations needed to estimate the expectation value of the original circuit increases polynomially with the number of tensor blocks \cite{guala2023practical}.  
Quantum convolutional neural networks (QCNNs) \cite{cong2019quantum} are among the variational quantum algorithms that utilize tensor-network-inspired and hierarchical architectures \cite{huggins2019towards,haghshenas2021optimization,
haghshenas2022variational}.  
While such structures often avoid barren plateaus, recent research suggests that architectures which provably do not exhibit barren plateaus can result in loss landscapes that are classically simulable using polynomial-time algorithms \cite{cerezo2023does,bermejo2024quantum}. This implies that although a quantum computer might be necessary for initial data collection and producing shadows of the input data,  a hybrid classical–quantum optimization loop is not required, and the parameterized quantum circuit need not be implemented on a quantum processor.
Further research is needed to determine whether non-classically simulable ansätze exist that are both practically useful for quantum machine learning and compatible with circuit cutting.

\subsection{Teleportation-based Methods}
\label{subsec:teleport}

\begin{figure}[t]
\includegraphics[width=\textwidth]{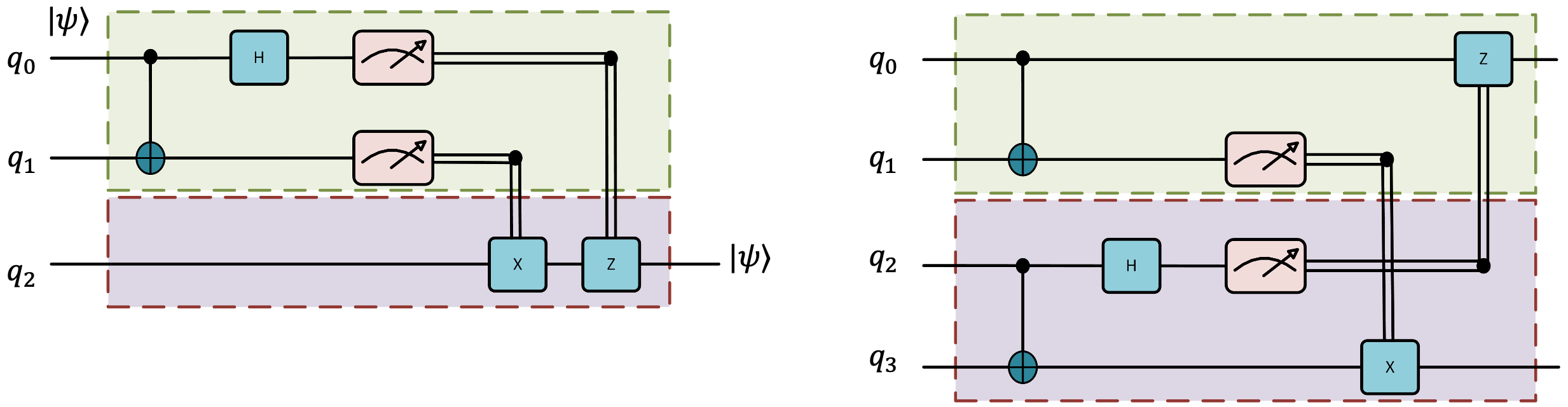}
\caption{The circuits on the left and right correspond to state teleportation and gate teleportation, respectively.  In state teleportation, Alice aims to transmit the quantum state $\rho = \ket{\psi}\bra{\psi}$ to Bob.  In contrast, the gate teleportation circuit shown here implements a controlled-NOT (CNOT) operation between two remote qubits $q_0$ and $q_3$ \cite{cuomo2020towards}.  In both circuits,  qubits $q_1$ and $q_2$ are prepared in the Bell state $\ket{\Phi^+}$, also known as an EPR pair \cite{einstein1935can}. }
\label{fig:tele}      
\end{figure}

Notable variants of quantum teleportation include state teleportation \cite{bennett1993teleporting,ren2017ground,cuomo2020towards,cacciapuoti2020entanglement}, entanglement swapping \cite{zukowski1993event,pan1998experimental}, and gate teleportation \cite{eisert2000optimal,huang2004experimental,meter2006architecture,jiang2007distributed}, which enable one-way  transfer of an unknown quantum state using shared entanglement and classical communication, bi-directional entanglement distribution, and remote gate application, respectively \cite{barral2024review}.  
Here, we only review state and gate teleportation. To learn more about other variants of quantum teleportation and their important applications in quantum technologies,  readers can refer to the reviews by Barral et al. \cite{barral2024review} and Horodecki et al. \cite{horodecki2009quantum}.

\subsubsection{State Teleportation}
\label{subsec:state-teleport}

Quantum state teleportation \cite{bennett1993teleporting,bouwmeester1997experimental,van2006distributed}  enables the transfer of an arbitrary quantum state $\rho$ between two parties using a single entangled qubit pair (an e-bit).  
In contrast to remote state preparation  \cite{bennett2001remote},  the quantum state being transferred is unknown to both the sender and the receiver.  
Consider a scenario in which Alice intends to transmit an arbitrary quantum state $\rho$ to Bob. Due to the constraints imposed by the no-cloning theorem \cite{wootters1982single,dieks1982communication,milonni1982photons,barnum1996noncommuting}, it is not possible for Alice to create and send a duplicate of the state.  Instead, she employs a quantum teleportation protocol.  Specifically, Alice conducts a Bell-state measurement (BSM) on the qubit representing the state $\rho$ and one half of an entangled pair that she shares with Bob.  This measurement projects her two qubits randomly into one of the four maximally entangled Bell states \cite{nielsen2010quantum}—
$\ket{\Phi^{\pm}}$ or $\ket{\Psi^{\pm}}$—
 each  with equal probability,  where
 \begin{align*}
\ket{\Phi^{\pm}} = \frac{1}{\sqrt{2}} (\ket{00} \pm \ket{11} ),
\ket{\Psi^{\pm}} = \frac{1}{\sqrt{2}} (\ket{01} \pm \ket{10} ).
 \end{align*}
Bob's qubit, which is entangled with Alice's, collapses into the state
$B^\dagger \rho B$,  where $B \in \{I,Z,X,ZX\}$ corresponds to the outcome of the BSM.
Alice then classically communicates the result of her measurement to Bob via two classical bits.  After getting the message, Bob applies the correct operation to his qubit to recover the original state $\rho$.

\subsubsection{Gate Teleportation}
\label{subsec:gate-teleport}

Gate teleportation \cite{eisert2000optimal,huang2004experimental,meter2006architecture,jiang2007distributed} allows the implementation of a two-qubit controlled unitary operation on unknown control and target states using a single entangled qubit pair, without physically transferring either qubit between two quantum processors.
Similar to state teleportation, gate teleportation also requires classical communication between the two parties to enable the application of appropriate corrections.

Unlike quantum circuit cutting, quantum teleportation does not introduce an exponential sampling overhead. However, teleportation consumes entangled pairs of qubits shared between parties. Minimizing the e-bit cost 
as well as optimizing partitioning strategies for distributed quantum circuits, has been an active area of research \cite{zomorodi2018optimizing,heunen2019automated,houshmand2020evolutionary,baker2020time,nikahd2021automated,ferrari2021compiler,
g2021efficient,cuomo2023optimized,ferrari2023modular,burt2024generalised}.

\section{Adversarial Robustness in Quantum Federated Learning}
\label{sec:adv-federated}

In this section, we begin by reviewing adversarial attacks that target classical federated learning architectures,  encompassing both privacy breaches and attempts to degrade or disrupt system performance.  This review,  presented in Section \ref{sec:adv-federated-overview},  is essential,  as quantum federated learning systems may also be susceptible to similar types of attacks.
While Figure \ref{fig:adv-fl} presents an overview of defense mechanisms applicable to classical federated learning systems,  we do not investigate them in depth.  Rather, our primary focus in Sections \ref{subsec:adv-federated-priv-leak} and \ref{subsec:adv-federated-integ} is on the defense strategies proposed for quantum federated learning.  Some of these defense methods are extensions of classical federated learning techniques adapted to the quantum realm, including differential privacy \cite{zhou2017differential,chen2024robust}, homomorphic encryption \cite{rivest1978data,fontaine2007survey,xu2023secure,chu2023cryptoqfl}, and secure multi-party computation \cite{yao1986generate,canetti2000security,li2024privacy} for privacy preservation, as well as adversarial training \cite{maouaki2025qfal} to enhance system robustness.  Other methods, however, are intrinsic to quantum computing itself, such as the inherent resilience of overparameterized variational quantum classifiers to adversarial attacks \cite{kumar2023expressive,heredge2024prospects,papadopoulos2025numerical},  and secure quantum protocols that rely on quantum communication, including quantum key distribution \cite{bennett1984quantum}, quantum secret sharing \cite{hillery1999quantum},   and blind quantum computing \cite{broadbent2009universal,li2021quantum,polacchi2023multi}.

\subsection{Overview of Adversarial Attacks in Classical Federated Learning}
\label{sec:adv-federated-overview}

Adversarial attacks in federated learning can be broadly categorized into privacy-leakage attacks and  integrity-oriented adversarial attacks,  with threats potentially originating from both the server and client sides \cite{zhou2022multi,zhou2021augmented,
zhang2024survey,wei2025trustworthy}.  
Privacy-leakage attacks include reconstruction and inference attacks \cite{zhang2024survey}.    
Reconstruction attacks aim to retrieve clients' datasets and include both gradient-based and parameter-based attacks \cite{zhang2024survey}.  Gradient-based attacks exploit shared gradients to extract original data samples \cite{zhu2019deep,zhao2020idlg}, while parameter-based attacks apply to scenarios where clients share model parameters with the server rather than gradients \cite{mcmahan2017communication,hitaj2017deep,wang2019beyond,xu2020information,yuan2021beyond}.
In contrast to reconstruction attacks, inference attacks—which include membership inference and property inference—aim to uncover properties of the data rather than reconstructing it \cite{zhang2024survey}.
Membership inference \cite{fredrikson2015model,shokri2017membership,lu2020sharing,li2021membership} aims to identify whether a specific sample was included in the training dataset, while property inference \cite{wang2019eavesdrop,melis2019exploiting,mo2020layer,xu2020subject,zhang2021leakage,luo2021feature} focuses on extracting key attributes from data that clients have not explicitly disclosed.  

\begin{figure}[p]
\includegraphics[width=0.9\textwidth]{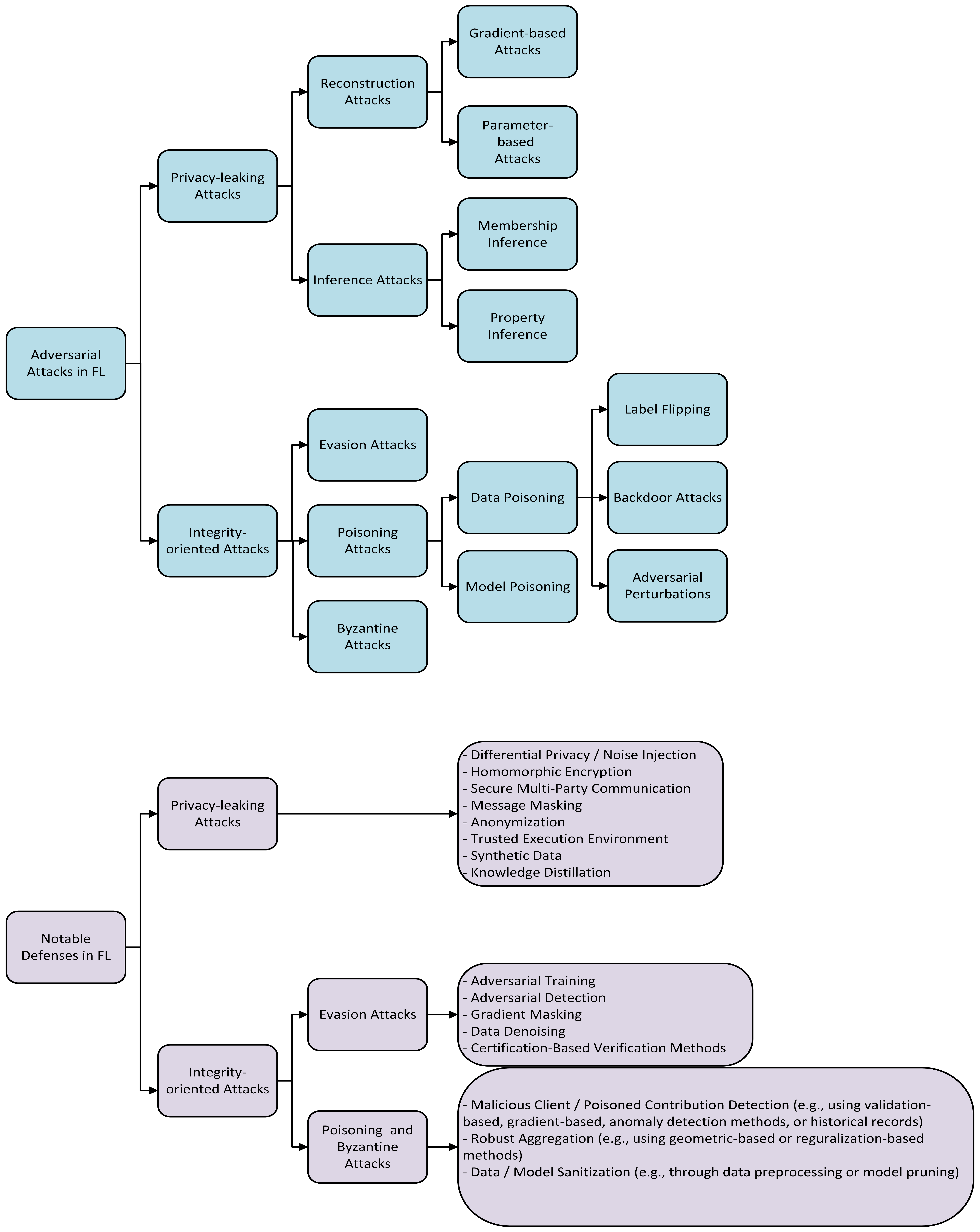}
\caption{This figure summarizes various categories of adversarial attacks in (classical) federated learning and notable defense methods. For further details on these attacks and the corresponding defense strategies,  refer to the relevant surveys on the subject \cite{kumar2023impact,zhang2024survey,wei2025trustworthy}.}
\label{fig:adv-fl}       
\end{figure}

Evasion, poisoning and Byzantine attacks 
are among the primary categories of integrity-oriented adversarial attacks in federated learning \cite{kumar2023impact,wei2025trustworthy}.  Evasion attacks involve generating adversarial examples by introducing small perturbations to input samples—typically preserving perceptual similarity for human observers—to deceive trained models into making incorrect predictions \cite{szegedy2013intriguing}.  
Poisoning attacks consist of data and model poisoning: the former involves tampering with training data to disrupt global model training, while the latter corrupts the model by altering the training process instead of the data itself \cite{wei2025trustworthy}.
Model poisoning could involve altering local model updates before sharing them with the server to manipulate the global model's outcomes \cite{jagielski2018manipulating,bhagoji2019analyzing,fang2020local}.  
Data poisoning encompasses both denial-of-service attacks that halt the learning process and more subtle strategies that target specific learning objectives while leaving most of the model's training unaffected  \cite{wei2025trustworthy},  often implemented via label flipping \cite{tolpegin2020data}, backdoor insertion \cite{xie2019dba,bagdasaryan2020backdoor}, or adversarial perturbations \cite{munoz2017towards,shafahi2018poison,zhu2019transferable,feng2019learning,koh2022stronger}.
Backdoor attacks \cite{gu2017badnets,li2020rethinking,bagdasaryan2020backdoor,khaddaj2023rethinking} subtly modify a small portion of the training data, usually by embedding a specific pattern known as a trigger into a chosen class.  
This allows attackers to manipulate the model’s predictions later by including the trigger in test-time inputs.  
In a federated learning setup, backdoor attacks can be carried out by corrupting the contributions of multiple clients, allowing the attacker to insert hidden behaviors into the shared model \cite{xie2019dba,bagdasaryan2020backdoor,
wang2020attack,mei2023privacy}.       
      Poisoning attacks based on adversarial perturbations involve tweaking training samples through gradient-based methods \cite{munoz2017towards}.  
 A Byzantine failure \cite{blanchard2017machine,lamport2019byzantine} in federated learning refers to a subset of nodes behaving arbitrarily or maliciously.  
 If the server aggregates these corrupted updates, the federated learning process could be disrupted \cite{zhang2024survey}. 
 Data and model poisoning attacks are sometimes regarded as a type of Byzantine failure,  while other Byzantine failures in federated learning could be a result of unreliable
communication and noisy data samples or models \cite{ang2020robust,fang2020local,tolpegin2020data,
rong2022poisoning,zhang2024survey}.  
 Figure \ref{fig:adv-fl} provides an overview of the different categories of adversarial attacks and key defense techniques in federated learning. 
For a more detailed examination of adversarial attacks in classical federated learning and potential defense strategies,  
readers can refer to several comprehensive surveys in this field \cite{almutairi2023federated,kumar2023impact,
xie2024survey,huang2024federated,
zhang2024survey,wei2025trustworthy}.

\subsection{Adversarial Attacks in Quantum Federated Learning}
\label{sec:adv-q-federated}

Adversarial attacks in quantum federated learning can be categorized similarly to those in classical federated learning,  although the methods for implementing these attacks may differ.
For instance, consider evasion attacks on quantum machine learning models in federated learning systems. These models may be trained on either native quantum data or classical data encoded into quantum states.
In classical machine learning,  adversarial examples are typically created by making small perturbations to the inputs of classifiers.  In quantum machine learning,  similar adversarial modifications can be introduced by either applying unitary perturbation operators to quantum input states or by perturbing classical inputs before they are encoded into quantum states \cite{lu2020quantum, gong2022universal,anil2024generating,kananian2025adversarial}.  To ensure that the perturbations remain small, such unitary  perturbation operators are often constrained to be close to the identity operator.

In the context of privacy-leaking attacks, one way they can be implemented in federated learning systems using quantum communication is through eavesdropping techniques,  such as the intercept-resend or Trojan horse attacks \cite{zhang2022federated}. 
In quantum communication, an intercept-resend attack occurs when an attacker intercepts the qubits in transit and then prepares and sends new qubits to the receiver
\cite{curty2005intercept,makarov2005faked,
deng2005improving,cai2006eavesdropping,
yang2013enhancement,lin2013intercept}.  
Trojan horse attacks 
\cite{vakhitov2001large,deng2005improving,
li2006improving,gisin2006trojan,luo2024security} 
 involve sending covert, unauthorized optical pulses into a legitimate quantum communication device. A portion of these Trojan signals becomes modulated with the legitimate information and is subsequently reflected back into the communication channel.  By examining the modulated reflections, the attacker can eavesdrop on the communication \cite{luo2024security}.
For a more comprehensive overview of various attacks in quantum communication,  readers may refer to the survey by Kumar et al. \cite{kumar2021state}.

\subsection{Resilience to Privacy-Leakage Attacks}
\label{subsec:adv-federated-priv-leak}

\subsubsection{Differential Privacy}
\label{subsec:adv-federated-priv-leak-dp}

Differential privacy (DP) \cite{dwork2006calibrating} is a mathematical framework for sharing 
aggregate statistics about
a dataset while limiting the amount of information leaked about specific individuals. Informally, an algorithm satisfies differential privacy if, upon observing its output, one cannot determine whether any individual’s data was included in the computation. Consequently, the behavior of a differentially private algorithm remains nearly unchanged when a single individual is added to or removed from the dataset. This guarantee applies to every individual, providing a formal guarantee that limits information leakage.  
The core idea of differential privacy is to add carefully calibrated noise to statistical outputs in a way that preserves the privacy of individuals.
Due to its relatively straightforward implementation and competitive computational and communication costs, differential privacy is one of the leading privacy-preservation mechanisms in machine learning \cite{pan2024differential,wei2025trustworthy},  especially in settings like federated learning.  It also has applications in other areas, such as distributed optimization \cite{huang2015differentially,huang2019dp} and mechanism design in game theory \cite{mcsherry2007mechanism,fain2016core,kananian2024asymptotically}.  
\begin{definition}[Differential Privacy \cite{dwork2006calibrating}]
A randomized mechanism $\mathcal{M}: \mathcal{D} \rightarrow \mathcal{R}$ satisfies $(\epsilon,\delta)-$differentially privacy if for any pair of adjacent input sets $X,X^\prime  \subseteq  \mathcal{D}$ and every subset of outputs $O \in \mathcal{R}$,  we have
\[
\mathbb{P}[\mathcal{M}(X) \in O]
\le e^\epsilon \mathbb{P}[\mathcal{M}(X^\prime) \in O] + \delta.
\]
\label{def:dp}
Here,  $\mathcal{D}$ and $ \mathcal{R}$ denote the domain and range of the mechanism, respectively,   while $\epsilon$ and $\delta$ control the intended privacy guarantees.  Generally,  smaller values yield stronger privacy guarantees,  though at the cost of introducing more noise, which can negatively impact output quality.  
\end{definition}
Differential privacy has been extended into the quantum realm \cite{zhou2017differential},  with various definitions of quantum differential privacy (QDP) emerging based on the choice of distance metrics used to define neighboring quantum states \cite{zhou2017differential,aaronson2019gentle,
hirche2023quantum,angrisani2023unifying,gong2024enhancing}.  
Here, we present the definition based on the trace distance \cite{zhou2017differential}.
For a discussion of alternative definitions of quantum differential privacy and the distance metrics they employ, see 
the survey by Zhao et al.  \cite{zhao2024bridging}.
\begin{definition}[Quantum Differential Privacy \cite{zhou2017differential}]
A quantum operation $\mathcal{E}$ satisfies $(\epsilon,\delta)-$differentially privacy if for any pair of adjacent inputs $\rho,\sigma$ such that $\tau(\rho,\sigma) \le d$,  and for every POVM $M  = \{M_m\}$ and all $O \subseteq \Omega(M)$,  we have
\[
\mathbb{P}[\mathcal{E}(\rho) \in_M O]
\le e^\epsilon \mathbb{P}[\mathcal{E}(\sigma) \in_M O] + \delta,
\]
\label{def:qdp}
where $\mathcal{E}$ and $\tau$  denote a completely positive and trace-preserving (CPTP) map and the trace distance,  respectively.  Furthermore,  $\rho$ and $\sigma$ are density operators,  $d\in (0, 1]$,  POVM stands for Positive Operator-Valued Measure,  and $\Omega(M) = \{m\}$ represents the set of all possible outcomes of $M$.
\end{definition}

Differential privacy has been shown to provide certified robustness against adversarial attacks in classical classifiers \cite{cohen2019certified}.  This connection between differential privacy and certified robustness has also been explored in the context of quantum machine learning 
\cite{weber2021optimal,du2021quantum,
huang2023certified,wu2023radio,winderl2023quantum,gong2024enhancing}.
The depolarization noise present in NISQ-era quantum classifiers can make them inherently quantum differentially private and naturally resilient to adversaries \cite{du2021quantum}.   
Differential privacy can be achieved in quantum algorithms through internal or external randomization mechanisms.  These mechanisms may influence the state preparation phase, quantum circuits,  or the measurement process \cite{zhao2024bridging}.  

Encoding classical information into quantum states can inherently yield $(\epsilon, \delta)$-differential privacy (Definition \ref{def:dp}) \cite{angrisani2022differential}.
Furthermore,  in quantum algorithms that use classical data encoded as quantum states, noise can be added to the classical data prior to encoding \cite{senekane2017privacy}.  Since the classical input to the algorithm satisfies $\epsilon$-differential privacy in this scenario, the quantum algorithm also satisfies $\epsilon$-differential privacy (Definition \ref{def:dp}),  by the post-processing property of differential privacy \cite{dwork2014algorithmic,zhou2017differential}. 
To achieve quantum differential privacy (Definition \ref{def:qdp}),  randomized encoding can be applied to quantum states \cite{du2022quantum,huang2023certified,gong2024enhancing}.
QDP and certified robustness against adversarial attacks can be attained by introducing quantum noise through random rotation gates applied to the input states of quantum classifiers \cite{huang2023certified}, or by randomly encoding the inputs using unitary transformations or quantum error correction encoders \cite{gong2024enhancing}.
In variational quantum classifiers \cite{cerezo2021variational},  where classical optimizers are used to tune the parameters of a parameterized quantum circuit,  one approach to achieving differential privacy is to introduce noise into the classical optimization process \cite{watkins2023quantum, rofougaran2024federated}.
Conversely, shot noise inherent in quantum measurements can naturally induce QDP in quantum algorithms \cite{li2024differential}.
Differentially private quantum algorithms can also be realized through the deliberate manipulation of quantum measurements \cite{angrisani2022quantum}.

Numerous studies have explored the role of differential privacy in quantum federated learning \cite{li2021quantum,xu2023secure,ullah2024quantum,
rofougaran2024federated,bhatia2024federated,
chen2024robust,moore2025quantum}.  
For instance,  
Rofougaran et al. \cite{rofougaran2024federated} propose an approach that incorporates differential privacy into the training of each local client by adding noise to the classical optimization process used to train the parameters of variational quantum classifiers. 
Bhatia et al. \cite{bhatia2024federated} also achieve local differential privacy by clipping random samples from clients' data and adding noise to the clipped gradients.
In contrast, Chen et al. \cite{chen2024robust} leverage quantum noise to mitigate privacy leakage and enhance robustness against adversarial attacks in quantum federated learning.

\subsubsection{Intrinsic Privacy}
\label{subsec:adv-federated-intrinsic}

In federated learning,  sharing gradients with a honest-but-curious server could potentially compromise clients' private information \cite{zhao2020idlg,huang2021evaluating,kumar2023expressive}.   
An honest-but-curious server carries out gradient aggregation but may also attempt a local gradient inversion attack to infer information about client data.
Kumar et al.  \cite{kumar2023expressive} show that in a federated learning environment where variational quantum circuits \cite{benedetti2019parameterized,cerezo2021variational} 
are used in place of classical neural networks,  highly expressive and overparametarized circuits provide a form of intrinsic protection
from gradient inversion attacks.
Here,  high expressiveness refers to the presence of a large number of distinct, non-degenerate Fourier frequencies when the circuit's output is represented in the Fourier domain \cite{gil2020input,schuld2021effect,
shin2023exponential,landman2022classically,
herman2023expressivity}.   
    Specifically,  Kumar et al.  \cite{kumar2023expressive} 
 explore overparameterized,  hardware-efficient ansätze \cite{moll2018quantum},  where the number of trainable parameters grows exponentially with the qubit count, alongside expressive encoding schemes that induce an exponential growth in the number of frequencies per input dimension when expressing the output and cost function gradients.
They demonstrate that performing gradient inversion to recover the data leads to solving systems of high-degree multivariate Chebyshev polynomial equations,  where the polynomial degree grows exponentially with the number of qubits.  
The number of such equations depends on the amount of shared gradient information, which is, in turn, tied to the number of trainable parameters.
Furthermore, they show that both the time and memory required to solve these equations—whether exactly or approximately—increase exponentially with the number of qubits.
Another approach to recovering client data involves machine learning-based methods that generate dummy gradients and train an attack model to replicate the client’s gradients \cite{zhao2020idlg,huang2021evaluating,eloul2022enhancing}. 
These attacks typically minimize the absolute distance between the dummy and real gradients by optimizing a dummy input vector designed to approximate the client’s original input \cite{zhu2019deep,geiping2020inverting,eloul2024mixing}—though various adaptations of this approach exist \cite{geiping2020inverting,yin2021see}.  
As shown by Kumar et al.  \cite{kumar2023expressive},  these methods also fail when the attack model is underparameterized and highly expressive.  This is primarily due to the attack model's loss landscape, which contains an exponential number of isolated local minima, making the model effectively untrainable.

One drawback of Kumar et al.'s \cite{kumar2023expressive} work is that overparameterization of the ansatz may result in barren plateau  problems \cite{mcclean2018barren} and hinder trainability \cite{heredge2024prospects}.  
The relationship between barren plateaus and adversarial robustness in variational quantum classifiers has also been explored by Gong et al. \cite{gong2024enhancing}, who show that adding randomized encoders to quantum circuits can lead to barren plateaus, obscuring gradient information from potential adversaries and impeding gradient-based adversarial algorithms in generating adversarial perturbations.  
A barren plateau refers to a region in the loss landscape where the loss values become exponentially concentrated as the problem size increases, with the loss gradients vanishing with high probability for randomly selected parameter values \cite{mcclean2018barren,larocca2024review}.  Along with poor local minima \cite{bittel2021training, anschuetz2021critical} and limited expressivity \cite{tikku2022circuit}, barren plateaus represent one of the three major obstacles to the trainability of variational quantum algorithms, and a significant body of work has been dedicated to studying them \cite{wang2021noise,cerezo2021higher,
arrasmith2022equivalence,qi2023barren,
cerezo2023does,larocca2024review,bermejo2024quantum}.  
In a recent study, Heredge et al. \cite{heredge2024prospects} theoretically investigate the trade-off between privacy protection and trainability in variational quantum classifiers, establishing a connection between privacy vulnerabilities in these models and the dimension of 
the Lie algebra of the generators of their circuits.

Despite the potential resilience of overparameterized ansätze against gradient inversion attacks \cite{kumar2023expressive}, which may, however, lead to barren plateau problems and hinder trainability \cite{heredge2024prospects,gong2024enhancing},   
 recent work by Papadopoulos et al. \cite{papadopoulos2025numerical} introduces an inversion attack capable of recovering private training data from variational quantum classifiers in a federated learning setting. 
This approach integrates adaptive low-pass filters into the Finite Difference Method (FDM) for numerically computing gradients, helping the optimization process find the global minimum when minimizing the absolute distance between dummy and real gradients, despite the presence of many local minima. This is achieved by tuning the filter’s window to suppress frequencies associated with local minima. 

\subsubsection{Secure Quantum Protocols}
\label{subsec:adv-federated-priv-Q-protocol}

Integrating quantum capabilities into federated learning enables secure and efficient communication via quantum channels, leveraging methods like quantum key distribution \cite{bennett2014quantum,mehic2020quantum}  and quantum secret sharing \cite{hillery1999quantum} to reduce reliance on resource-intensive encryption \cite{sheng2017distributed,  zhang2022federated}. 
Quantum Key Distribution (QKD) \cite{bennett1984quantum,shor2000simple,bennett2014quantum,mehic2020quantum} is a cryptographic method that leverages the principles of quantum mechanics—such as the no-cloning theorem \cite{wootters1982single,dieks1982communication,milonni1982photons,barnum1996noncommuting}— 
to enable two parties to generate and share a secret key with information-theoretic security. It ensures that any attempt at eavesdropping introduces detectable disturbances in the quantum states, allowing the communicating parties to identify potential security breaches.
Numerous studies \cite{xu2022privacy,kaewpuang2023adaptive,park2023hqk,ren2024variational,tanbhir2025quantum} have explored the application of quantum communication and QKD in federated learning, including those that focus on the allocation of quantum communication resources (such as key generation rates and QKD links) and the routing of data or key material across the network \cite{xu2022privacy,kaewpuang2023adaptive,ren2024variational}.  
Building upon the  quantum secret sharing protocol \cite{hillery1999quantum},  Zhang et al. \cite{zhang2022federated} present a secure aggregation framework based on GHZ states \cite{greenberger1989going} that is  applicable to both classical and federated learning with conventional models such as neural networks and quantum federated learning employing variational quantum circuits.   This framework ensures security against both external eavesdroppers seeking to infer private information and internal semi-honest participants—those who follow the protocol correctly but attempt to covertly extract sensitive data.  
Assuming malicious participants do not collude, the physical properties of quantum communication enable the detection of both external eavesdropping and internal attacks  
  \cite{bennett2014quantum,shor2000simple}.

An additional method that can be incorporated into federated learning via quantum communication is blind quantum computing (BQC) \cite{aaronson2017complexity,polacchi2023multi}.  This technique allows clients to offload quantum computations to an untrusted server while keeping their data 
and algorithms confidential \cite{qu2021secure,chehimi2023foundations}.
Li et al. \cite{li2021quantum} propose a method for federated and private distributed learning based on the universal blind quantum computation protocol (UBQC) introduced by Broadbent et al.  \cite{broadbent2009universal}.   
This blind quantum computation protocol offers unconditional security without relying on computational assumptions,  enabling a client to offload a quantum computation to a server without disclosing any information about the computation, including its inputs or outputs. The client requires neither quantum memory nor significant computational power—only the ability to prepare qubits randomly selected from a finite set. The server performs the computation by receiving these qubits and following measurement instructions sent by the client via classical communication.
In the protocol proposed by Li et al.  \cite{li2021quantum},  
clients employ the UBQC protocol to outsource their gradient computations to the server.  
To ensure differential privacy and mitigating the risk of gradient inversion attacks by potential eavesdroppers during model training,  noise is added to the computed gradients before uploading them.  
This work achieves differential privacy through classical, not quantum, noise.

\subsubsection{Homomorphic Encryption}
\label{subsec:adv-federated-priv-leak-QHE}

Homomorphic Encryption (HE) \cite{rivest1978data,fontaine2007survey,gentry2009fully-book} is a privacy-preserving technique that allows computations to be performed directly on ciphertexts without decrypting it. HE produces an encrypted result that, when decrypted, matches the result of operations performed on the plaintext.  The applications of HE in classical federated learning have been extensively studied, especially in the context of reducing the encryption and communication overhead it introduces \cite{aono2017privacy,xu2019verifynet,
zhang2020batchcrypt,jin2023fedml,qiu2024hashvfl}.  
More recently,  HE has also been explored in the context of quantum federated learning \cite{xu2023secure,chu2023cryptoqfl,li2024quantum,dutta2024federated,dutta2024mqfl}.  For instance,  Xu et al. \cite{xu2023secure} apply local differential privacy and homomorphic encryption in a federated learning framework for autonomous vehicular networks that incorporates quantum communication. In their simulations, the system employs classical convolutional neural networks as the underlying machine learning model.  
Quantum Homomorphic Encryption (QHE) \cite{broadbent2015quantum,dulek2016quantum,mahadev2020classical} is the quantum counterpart to classical homomorphic encryption, allowing quantum circuits to be evaluated on encrypted quantum data. Chu et al. \cite{chu2023cryptoqfl} integrate QHE into a federated learning framework that leverages quantum communication for training quantum neural networks (QNNs) \cite{schuld2014quest,dunjko2018machine,jeswal2019recent,
cong2019quantum,perez2020data,abbas2021power}.  
In their approach, gradients are encoded into quantum states and homomorphically aggregated using quantum adders \cite{kumar2017optimal,sarma2018quantum}.

On the other hand, some encryption methods are vulnerable to attacks by quantum computers, as a malicious attacker equipped with quantum capabilities could potentially access plaintext data. Consequently,  
 there has been increasing interest in developing secure federated learning systems that are resilient to quantum threats, including the advancement of homomorphic encryption techniques designed to safeguard against these attacks \cite{yang2022post,zuo2023post,gurung2023secure,yu2024hqsfl,zhang2024pqsf,gharavi2025pqbfl,zhang2025efficient,appiah2025enhanced,qin2025efficient,commey2025pqs}.

\subsubsection{Secure Multi-Party Computation}
\label{subsec:adv-federated-priv-leak-QSMPC}

Secure Multi-Party Computation (SMPC) \cite{yao1986generate,canetti2000security,
cramer2015secure,goldreich2019play}  
allows multiple participants to collaboratively compute a function over their private data without revealing that data to each other or to a central server. No party learns anything beyond their own input and the final result they are meant to receive.
SMPC can be applied in federated learning to protect interactions between parties without disclosing private data \cite{ravikumar2023quantum}.  
For example, it can be used to securely aggregate gradients during model training \cite{yu2022quantum}.

A key benefit of integrating quantum capabilities into federated learning is the ability to efficiently encode classical data using a logarithmic number of qubits.
By encoding model contributions into quantum states, Li et al. \cite{li2024privacy} propose two frameworks for secure model aggregation in federated learning systems that leverage quantum communication: one utilizing private inner product estimation and the other implementing incremental learning.  
Incremental learning refers to the process where a model learns continuously, incorporating new data or updates over time without retraining from scratch.
In the first protocol,  secure model aggregation is reformulated as a correlation estimation task, allowing the blind quantum bipartite correlator (BQBC) algorithm \cite{li2024blind} to be adapted for use in a multi-party setting.
Built on quantum counting \cite{brassard1998quantum,tang2023communication},  the BQBC algorithm facilitates blind quantum machine learning \cite{zhou2021blind} by enabling inner product estimation, a fundamental operation in many widely used machine learning methods.
In the second protocol, clients engage in multi-party computation without server participation until the final stage, where the server retrieves the aggregated gradient information.
By using quantum communication, these privacy-preserving mechanisms achieve reduced communication costs compared to classical approaches.

\subsection{Robustness to Integrity-Oriented Attacks}
\label{subsec:adv-federated-integ}

\subsubsection{Byzantine Attacks}
\label{subsec:adv-federated-integ-byz}

Byzantine faults \cite{lamport2019byzantine,castro1999practical,driscoll2004real} 
refer to failures in a distributed system where components may act arbitrarily or maliciously, including lying or sending conflicting information to different parts of the system.  
Several Byzantine-tolerant algorithms have been proposed for applications in classical distributed machine learning \cite{blanchard2017machine,damaskinos2018asynchronous,xie2018generalized,yin2018byzantine,chen2017distributed,alistarh2018byzantine,xia2019faba,xia2020defenses,xie2020zeno++,xie2019zeno}.  
Xia et al.  \cite{xia2021defending} analyze the differences in Byzantine problems between classical  distributed learning and quantum federated learning,  and adapt several Byzantine-resilient algorithms
\cite{blanchard2017machine,xia2019faba,xia2021tofi}  
 designed for classical distributed learning to fit the quantum federated learning framework proposed in their earlier work 
 \cite{xia2021quantumfed}.  
 These Byzantine-robust algorithms include Krum \cite{blanchard2017machine},  FABA \cite{xia2019faba},  and ToFi \cite{xia2021tofi}.  
Krum and FABA are geometric-based approaches \cite{zhang2024survey} that assume malicious updates lie far from benign ones in terms of distance.
ToFi operates based on a reference dataset. It evaluates the loss on this dataset for each weight update uploaded by participants. Updates that lead to high losses are assumed to originate from Byzantine nodes and are removed.

\subsubsection{Evasion and Poisoning Attacks}
\label{subsec:adv-federated-integ-evasion}

Similar to their classical counterparts, quantum federated learning systems are vulnerable to evasion attacks, such as adversarial input perturbations \cite{maouaki2025qfal},  as well as poisoning attacks, such as label-flipping \cite{bhatia2024robustness,bhatia2024robustness2}.
Bhatia et al.  \cite{bhatia2024robustness,bhatia2024robustness2} investigate the robustness of a quantum federated learning system, using variational quantum classifiers for lithography hotspot detection, against label-flipping attacks,  which may result from either malicious intent or human error. Hotspots refer to specific areas on the wafer in semiconductor manufacturing that signal potential irregularities or defects during production, often arising from challenges associated with the printability of particular layout patterns \cite{liebmann2006reducing}. 
To counter label-flipping attacks,  Bhatia et al. \cite{bhatia2024robustness2} propose a defense strategy that detects such attacks by analyzing pairwise Euclidean distances between clients, where large distances indicate that a specific client's update is adversarial and detrimental to training the global model.
In a related work, Lee et al.  \cite{lee2025auction} propose an auction-based approach to filter out untrustworthy clients in federated learning systems utilizing quantum neural networks. However, their work primarily focuses on mitigating the non-IID data distribution issues in federated learning, rather than addressing poisoning or Byzantine attacks.
Client selection and outlier exclusion in quantum federated learning have also been explored by Son and Park \cite{son2024toward},  who propose an approach based on assessing class imbalance in local models using entropy, as well as measuring quantum state dissimilarity between the global target model and individual local models.
Ma et al. \cite{ma2025robust} propose a decentralized framework for quantum kernel learning that employs a clipping-based aggregation mechanism to mitigate the impact of corrupted updates from faulty or adversarial nodes. In this approach, client data is clipped prior to aggregation.

To enhance resilience against adversarial input perturbations, 
Maouaki et al.  \cite{maouaki2025qfal} present a framework where the clients' variational quantum classifiers are adversarially trained within a quantum federated learning system.  For adversarial training, they leverage adversarial examples generated using PGD-based methods \cite{madry2017towards}.
Their results demonstrate that adversarial training, even when applied to only a subset of clients, enhances robustness against adversarial attacks.   In fact, varying the number of adversarially trained clients, as well as the strength of the perturbation, reveals a trade-off between accuracy on clean data and resilience to adversarial attacks.

\begin{figure}[htbp]
\includegraphics[width=\textwidth]{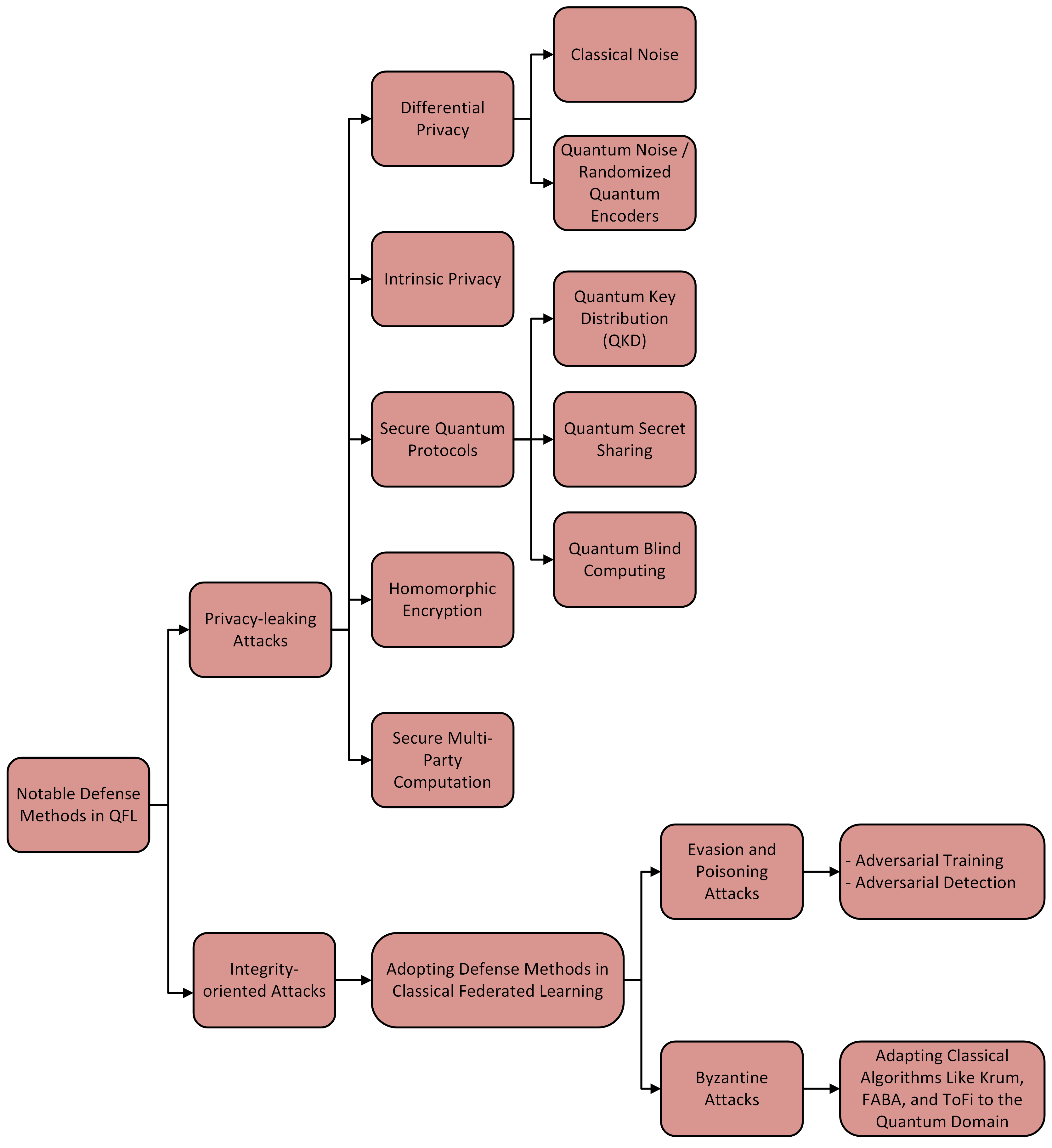}
\caption{An overview of the defense methods discussed in Sections \ref{subsec:adv-federated-priv-leak} and \ref{subsec:adv-federated-integ}. 
There is relatively less research on the adversarial robustness of quantum federated learning systems against integrity-based attacks compared to privacy-leaking attacks, and the defense methods proposed to date are typically built upon classical methods.}
\label{fig:def-qfl}       
\end{figure}

\section{Circuit cutting and Adversarial Robustness}
\label{sec:adv-cut}

The adversarial robustness of quantum classifiers subjected to circuit cutting has recently been studied \cite{kananian2025adversarial}. 
When quantum communication is unavailable, circuit cutting can be used to distribute the execution of a quantum circuit across multiple quantum processors.  However, as with other distribution methods, partitioning these circuits may increase their susceptibility to adversarial attacks.
The sub-circuits produced by circuit cutting can be executed in a distributed manner across multiple devices.  
If an adversary gains access to any of these sub-circuits, they could attempt to infer private information or launch various attacks to disrupt the system's functionality.
When the outputs of these sub-circuits are combined to reconstruct the original circuit's outcome,  
any manipulation of one or more sub-circuits by an adversary can cause the reconstructed result to differ from the intended one.  

\begin{figure}[t]
    \centering
    \includegraphics[width=\textwidth]{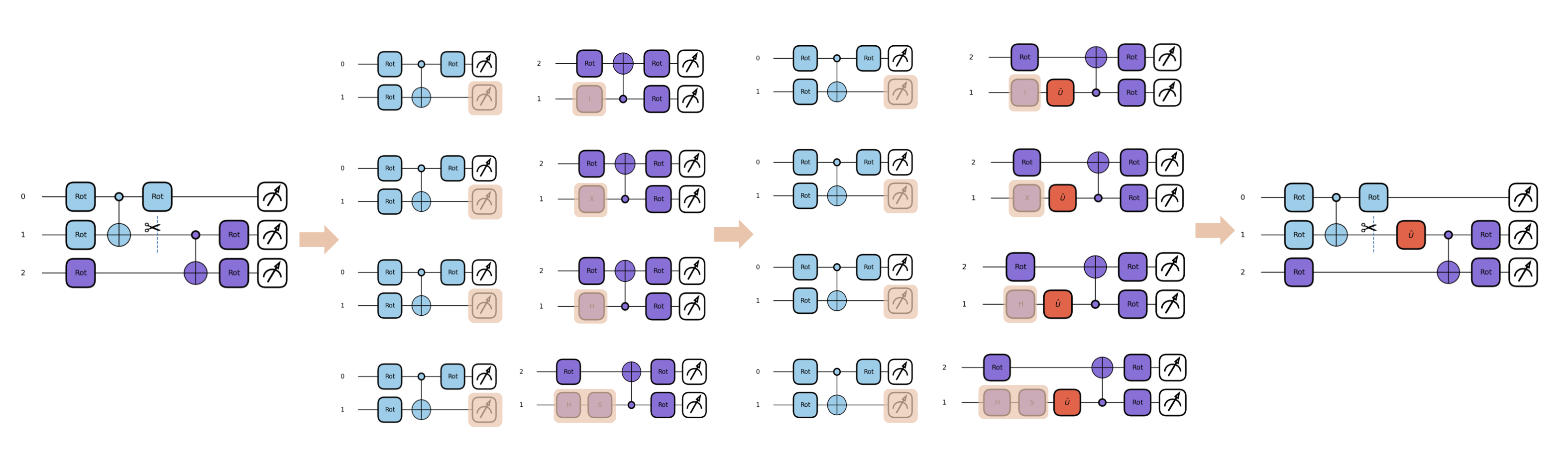}
    \caption{This figure,  drawn with the aid of Pennylane \cite{bergholm2018pennylane},  illustrates a simple quantum circuit undergoing wire cutting, resulting in multiple sub-circuits.  
    The sub-circuits shown in purple are then attacked by applying an adversarial perturbation gate $\hat{U}$ to their input states.  After combining the results of these sub-circuits to reconstruct the original circuit, this adversarial attack results in implementing an adversarial gate $\hat{U}$ within intermediate layers of the reconstructed circuit.  Wire cutting involves replacing identity channels (i.e., wires) with a linear combination of measurement and state preparation channels.  These measurement and state preparation operators are highlighted in the sub-circuits using cream-colored boxes. }
    \label{fig:adv-cutting}
\end{figure}

One possible method of attacking the sub-circuits is through evasion attacks and adversarial perturbations. 
Adversarial perturbations typically refer to slight modifications made to the input data of classifiers to trick the models into producing incorrect predictions \cite{szegedy2013intriguing}.  
In quantum classifiers, adversarial perturbations can be introduced either by applying perturbation gates to the quantum input states or by altering the classical inputs prior to their encoding into quantum states (see Section \ref{sec:adv-q-federated}).
When a circuit is partitioned by wire cutting, the input states of the resulting sub-circuits are either inherited from the original circuit's inputs or prepared specifically as part of the wire-cutting process.
As shown in Figure \ref{fig:adv-cutting},  when an adversary adds adversarial perturbation gates to the input states generated through wire cutting,  this modification leads to the implementation of an adversarial gate within intermediate layers of the reconstructed quantum circuit.  In a recent work \cite{kananian2025adversarial},  Kananian and Jacobsen theoretically and experimentally study the implications of such an attack.

The attacks illustrated in Figure  \ref{fig:adv-cutting} for wire cutting could be extended to a scenario where the quantum circuit is partitioned through state teleportation 
instead of wire cutting.  
An adversary with physical access to a sub-circuit at the receiving end of the quantum teleportation protocol can add perturbations to its received input state,  resulting in the introduction of an adversarial gate within the original circuit's layers.  
Alternatively,  a malicious node executing a sender sub-circuit can adversarially perturb a state prior to teleporting it to another sub-circuit.
Beyond evasion attacks,  future research should investigate other potential attack scenarios targeting partitioned quantum classifiers and possible methods for defending against them.

\section{Conclusions}
\label{sec:open-q}

This work has reviewed the current literature on adversarial robustness in quantum federated learning and partitioned quantum classifiers.
Both paradigms are important in distributed quantum machine learning. Studying the adversarial robustness of quantum models when deployed using these paradigms provides valuable insights into the potential advantages of QML models over classical ones,  and contributes to the design of systems that are resilient and trustworthy.

There remains significant opportunity to explore a broader range of attack and defense scenarios in quantum federated learning, particularly regarding integrity-oriented attacks, which have been less studied compared to privacy-leaking attacks.  While applying circuit distribution methods to quantum machine learning is increasingly important, especially in light of current limitations in quantum hardware, the adversarial robustness of partitioned quantum models has only recently begun to receive attention in the literature \cite{kananian2025adversarial}.  It is therefore essential to conduct a deeper investigation into the robustness of these systems and to develop effective defense mechanisms tailored to their unique characteristics.

\bibliographystyle{plain}
\bibliography{ref-chapter}

\end{document}